\documentclass{nature}
\usepackage{graphicx}
\usepackage{amsmath}%
\usepackage{amsfonts}%
\usepackage{amssymb}%
\usepackage{graphicx}
\usepackage{braket}
\usepackage{bbold}
\usepackage{textcomp}
\usepackage{xspace}
\usepackage{subfigure}
\usepackage[usenames,dvipsnames]{color}
\usepackage{hyperref}
\usepackage{moreverb}
\usepackage{color}
\usepackage{lineno}

\def\ket#1{\left|#1\right>}

\title{Spin-orbit coupled fermions in an optical lattice clock}

\author{S. Kolkowitz$^{1,*}$,
S.L. Bromley$^{1,*}$,
T. Bothwell$^{1}$,
M.L. Wall$^{1}$,
G.E. Marti$^{1}$,
A.P. Koller$^{1}$,
X. Zhang$^{1,\dagger}$,
A.M. Rey$^{1}$,
J. Ye$^{1}$\\
$^*$ These authors contributed equally.}

\begin{document}
\maketitle
\begin{affiliations}
 \item JILA, NIST and Department of Physics, University of Colorado, 440 UCB, Boulder, Colorado 80309, USA \\
 $^{\dagger}$Present address: International Center for Quantum Materials, School of Physics, Peking University, Beijing 100871, China
\end{affiliations}
\date{\today}

\noindent\textbf{Engineered spin-orbit coupling (SOC) in cold atom systems can aid in the study of novel synthetic materials and complex condensed matter phenomena\cite{DalibardReview,SpielmanReview,ManyBodyReview,galitskiSOCreview,zhaiSOCreview,SpielmanSOC,FallaniSOC,BlochChiral}. Despite great advances, alkali atom SOC systems are hindered by heating from spontaneous emission, which limits the observation of many-body effects\cite{DalibardReview,SpielmanReview,zhaiSOCreview}, motivating research into potential alternatives\cite{LevSOC,GyuBoong,KetterleSpinTex}. Here we demonstrate that SOC can be engineered to occur naturally in a one-dimensional fermionic $^{87}$Sr optical lattice clock (OLC)\cite{WallSOC}. In contrast to previous SOC experiments\cite{DalibardReview,SpielmanReview,ManyBodyReview,galitskiSOCreview,zhaiSOCreview,SpielmanSOC,FallaniSOC,BlochChiral,LevSOC,GyuBoong,KetterleSpinTex}, in this work the SOC is both generated and probed using a direct ultra-narrow optical clock transition between two electronic orbital states. We use clock spectroscopy to prepare lattice band populations, internal electronic states, and quasimomenta, as well as to produce SOC dynamics. The exceptionally long lifetime of the excited clock state (160~s) eliminates decoherence and atom loss from spontaneous emission at all relevant experimental timescales, allowing subsequent momentum- and spin-resolved \textit{in situ} probing of the SOC band structure and eigenstates. We utilize these capabilities to study Bloch oscillations, spin-momentum locking, and Van Hove singularities in the transition density of states. Our results lay the groundwork for the use of OLCs to probe novel SOC phases of matter.}

\begin{figure}
\begin{center}
\includegraphics[width=5.5in]{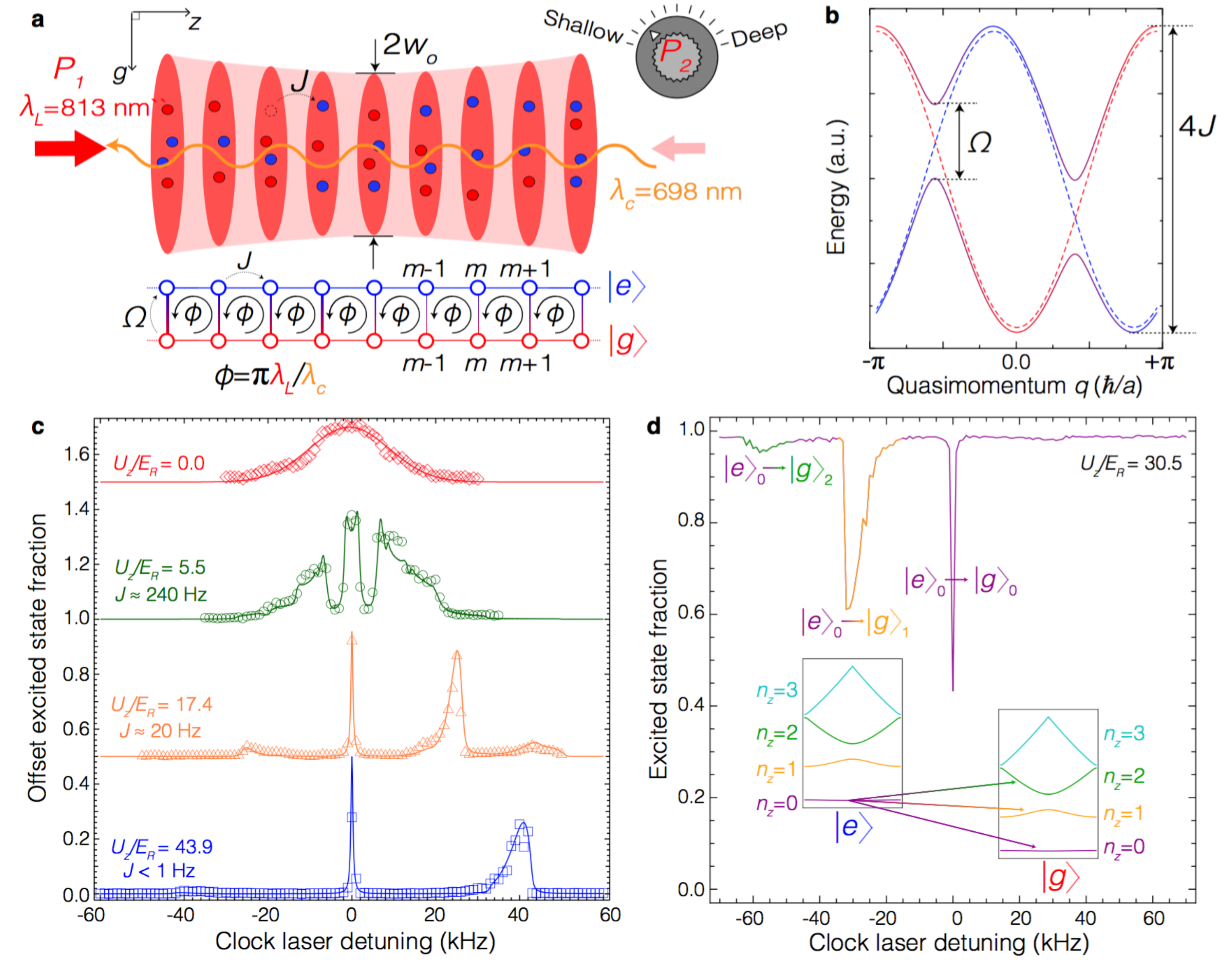}
\end{center}
 \caption{\textbf{Spin-orbit coupling (SOC) in a 1D lattice with tunable tunneling.} \textbf{a,} Atoms are trapped in a 1D optical lattice formed by interfering a strong trapping beam (power $P_{1}$, wavelength $\lambda_L=813$~nm) with a counter-propagating, tunably attenuated retro-reflection (variable power $P_{2}$, represented by a cartoon knob). The atoms are probed on the narrow clock transition with an ultra-stable clock laser ($\lambda_c=698$~nm, Rabi frequency $\Omega$). The resulting SOC Hamiltonian is equivalent to that of charged fermions on a synthetic 2D ladder, with horizontal tunneling rate $J$, vertical tunneling rate $\Omega$, and a synthetic magnetic field flux $\phi=\pi\lambda_L/\lambda_c$. \textbf{b,} The clock laser couples the dispersion curve for $\ket{g}_{n_z=0}$ (dashed red line) to the quasimomentum-shifted curve for $\ket{e}_0$ (dashed blue line), resulting in spin-orbit coupled bands (solid bichromatic lines). \textbf{c,} Clock spectroscopy ($\Omega\approx2\pi\times200$~Hz, 80 ms pulse duration) at four axial trapping potentials (data and fits are shifted upward for clarity). When $P_{2} = 0$ mW, the sidebands and carrier merge into a Doppler broadened Gaussian (red diamonds). The solids lines are theoretical calculations using a model that perturbatively treats the axial and radial coupling (see Methods). \textbf{d,} Spectroscopy of atoms in $\ket{e}_0$, prepared by driving the $\ket{g}_0\to\ket{e}_0$ transition, then removing any remaining atoms in $\ket{g}$.}
\label{fig_1}%
\end{figure}

When tunneling is allowed, spin-orbit coupling emerges naturally in a nuclear spin polarized $^{87}$Sr OLC during laser interrogation of the narrow linewidth ${}^{1}S_0 (\ket{g})-{}^{3}P_0 (\ket{e})$ clock transition at wavelength $\lambda_c=698$~nm (Fig.~\ref{fig_1}a). The lattice used to confine the atoms has a wavelength $\lambda_L=813$~nm. At this ``magic" wavelength, the band structures of the two clock states are identical with band energies $E_{n_z}(q)$, determined by the discrete band index $n_z$ and quasimomentum $q$ in units of $\hbar/a$, where the lattice constant $a=\lambda_L/2$ and $\hbar$ is the Planck constant divided by $2\pi$. When an atom is excited from $\ket{g}$ to $\ket{e}$ using a clock laser with Rabi frequency $\Omega$ and frequency detuning $\delta$ from the clock transition, energy and momentum conservation require a change in atomic momentum by $2\pi \hbar/\lambda_{c}$.

The resulting Hamiltonian can be diagonalized in quasimomentum space by performing a gauge transformation $\ket{e,q}_{n_z}\rightarrow \ket{e,q+\phi}_{n_z}$, where $\phi=\pi\lambda_L/\lambda_c\approx7\pi/6$. Fig.~\ref{fig_1}b shows the transformed $\ket{g}$ and $\ket{e}$ bands for $n_z=0$, the ground state band, under the rotating-wave approximation when $\delta=0$. The transformed SOC Hamiltonian is given by\cite{WallSOC}
 \begin{linenomath*}
\begin{align}
\label{eq:Hamilton}H_{SOC}&=-\hbar\sum_{q} \vec{B}_{n_{z}}(q,\Omega,\delta) \cdot \vec{S},
\end{align}
\end{linenomath*}
\noindent where the components of $\vec{S}$ are $\hat S^{X,Y,Z},$ the spin-1/2 angular momentum operators for the two clock states. $\vec{B}_{n_{z}}(q,\Omega,\delta)$ is an effective, quasimomentum-dependent magnetic field given by
 \begin{linenomath*}
\begin{align}
\label{eq:BlochVector} \vec{B}_{n_{z}}(q,\Omega,\delta) &=\left[B_{n_z}^X(\Omega),0,B_{n_z}^Z(q,\delta)\right]=\left[\Omega,0,(E_{n_z}(q)-E_{n_z}(q+\phi))/\hbar+\delta \right],
\end{align}
\end{linenomath*}
where in the tight binding limit $E_0(q)=-2\hbar J\cos(q)$, and $J$ is the tunneling rate between nearest neighbor lattice sites. The eigenstates of $H_{SOC}$ are described by Bloch vectors in the $\hat{X}\text{-}\hat{Z}$ plane, pointing along the magnetic field $\vec{B}_{n_z}(q,\Omega,\delta)$, with their orientation specified by the chiral Bloch vector angle $\theta_B$ with respect to the $\hat{Z}$ axis\cite{WallSOC}, where 
 \begin{linenomath*}
\begin{align}
\label{eq:theta_B} \theta_B=\arctan(\frac{\Omega}{(E_{n_z}(q)-E_{n_z}(q+\phi))/\hbar+\delta}). 
\end{align}
\end{linenomath*}
\noindent The $q$ dependence of $\theta_B$ is a manifestation of chiral spin-momentum locking\cite{WallSOC, ChiralLadders}.

To connect this system to related works on synthetic gauge fields\cite{SpielmanSOC,FallaniSOC,DalibardReview,SpielmanReview,BlochChiral,aidelsburger,ShakenLattice2,KetterleSpinTex}, we can treat the internal clock transition ($\ket{g}\to\ket{e}$) as a synthetic dimension\cite{LewensteinReview}, as shown in Fig.~\ref{fig_1}a. In this case, an atom following a closed trajectory about a single plaquette ($\ket{m,g}\to\ket{m+1,g}\to\ket{m+1,e}\to\ket{m,e}\to\ket{m,g}$) accumulates a phase, given by the same $\phi$ defined previously, which resembles the flux experienced by a charged particle in the presence of an external magnetic field. In this picture, the chiral Bloch vector angle $\theta_B$ is directly connected to the topological nature of chiral edge modes of the two-dimensional Hofstadter model\cite{ChiralLadders,BlochChiral}. Coupling multiple nuclear spin states with our synthetic gauge fields should enable the realization of topological bands\cite{DalibardReview,SpielmanReview,zhaiSOCreview} and exotic phases in higher dimensions\cite{MagneticXtals}.

In our experiment, several thousand nuclear-spin polarized fermionic $^{87}$Sr atoms are cooled and loaded into a horizontal one-dimensional optical lattice aligned along the $\hat{z}$-axis with $\sim2$~$\mu$K temperatures. The lattice is formed using a high power ($P_{1}\approx$~3 W) incoming beam focused down to a beam waist $w_{0}\approx45~\mu$m and a strongly attenuated retro-reflection with tunable power ($0\le P_{2}\le$~50~mW, Fig.~\ref{fig_1}a). This enables the radial trap frequency to effectively remain constant at $\nu_r \approx450$ Hz, while $U_{z}$ can be tuned via $P_{2}$ over a wide range from $U_{z}/E_{r}=0$ to $U_{z}/E_{r}>200$, where $E_{r}=\frac{\hbar^2 k_{L}^2}{2m}$ is the lattice recoil energy, and $m$ is the atomic mass. This corresponds to axial trap frequencies $\nu_z\approx2E_{r}\sqrt{U_{z}/E_{r}}/2\pi\hbar$ up to $\sim100$~kHz. When $\nu_z\gtrsim40$~kHz, site to site tunneling takes longer than experimentally relevant timescales and the atoms are effectively localized to single lattice sites, as is standard in OLC operation\cite{ClockSr}. However, for smaller $\nu_z$ tunneling between nearest neighbor lattice sites is important and occurs at a rate $J_{\mathbf{n_r}}$ that depends on the radial mode index $\mathbf{n_r}$. In this regime, atomic motion in the axial direction is described by delocalized Bloch states characterized by $n_z$ and $q$. For a $\sim2$~$\mu$K thermal distribution, the atoms are predominantly in the $n_z=0$ axial ground band and completely fill the band. The average radial mode occupation is $\langle{n_r}\rangle\sim$100.

The clock laser is locked to an ultra-stable optical cavity\cite{ClockSr} with a linewidth of $\sim$26 mHz. Because the clock laser is co-linear with the lattice axis, coupling to the radial motional modes is suppressed, and for the entirety of this work the system will be treated as quasi-one-dimensional, with relatively minor corrections arising from the thermal average of the Rabi frequency $\Omega_{\mathbf{n_r}}$ and the tunneling rate $J_{\mathbf{n_r}}$ over the radial mode occupation. For the sake of clarity we therefore drop the radial mode index from $\Omega$ and $J$. Throughout this work the clock laser Rabi frequency $\Omega$ is measured on resonance with the carrier at $\delta=0$ with a high axial trapping frequency $\nu_z>50$~kHz.  The mean particle number per lattice site was kept in the range $N\sim1-10$, which for the operating conditions results in a density-dependent many-body interaction rate of $N\chi/(2\pi)\lesssim1$~Hz, where $\chi$ is the two-body interaction rate\cite{ManyBodySr}. For the experiments presented here $\Omega\gg N\chi$, and thus the results are all well described by a single particle model.

Unlike previous studies of SOC in ultra-cold atoms in which time of flight (TOF) measurements are used to determine the momentum distribution\cite{DalibardReview,SpielmanReview,ManyBodyReview,galitskiSOCreview,zhaiSOCreview,SpielmanSOC,FallaniSOC,BlochChiral,LevSOC,GyuBoong,KetterleSpinTex}, all of the data presented in this work is measured \textit{in situ} using clock spectroscopy\cite{ClockSr}. Clock spectroscopy provides precise measurement and control of the atomic spin and motional degrees of freedom, access to the atomic density of states, and offers the prospect for real-time, non-destructive measurement of atom dynamics in the lattice. At the end of each experiment the number of atoms in the $\ket{e}$ and $\ket{g}$ states are counted using a cycling transition (see Methods), and the normalized population fraction in each state is extracted. For example, Fig.~\ref{fig_1}c presents spectroscopy of the carrier and motional sideband transitions at four different axial trapping potentials, with the atoms initially prepared in $\ket{g}_0$. Here each data point was taken with a new sample of $^{87}$Sr at a different clock laser detuning. At $U_{z}/E_{r}=43.9$ (blue squares) the atoms are strongly confined and the data is well described by a simple model that neglects tunneling between lattice sites\cite{Blatt}. However, as $P_{2}$ is turned down and the trapping potential is reduced to $U_{z}/E_{r}=5.5$ (green circles) the carrier transition exhibits a broad, sharp splitting, which is no longer consistent with atoms localized to single sites. A model that perturbatively treats the axial and radial coupling\cite{Blatt} fully reproduces the measured lineshapes (solid lines, see Methods).

The narrow $\ket{g}_0\to\ket{e}_0$ carrier transition centered at $\delta/(2\pi)=0$~kHz enables the preparation of atoms in the $\ket{e}_0$ state, from which spectroscopy can also be performed, as shown in Fig.~\ref{fig_1}d. Due to the long lifetime of $\ket{e}$ we do not observe spin state relaxation to $\ket{g}$ for the time scales explored in this experiment ($<150$ ms, see Methods). In addition to the carrier transition, motional sidebands corresponding to axial inter-band transitions are also apparent in Fig.~\ref{fig_1}c. The measured lineshape in this case is also fully captured by the perturbative model (solid lines, see Methods). At high trapping potential (blue squares) the prominent blue-detuned sideband at $\delta/(2\pi)\approx40$~kHz corresponds to the $\ket{g}_0\to\ket{e}_1$ transition. The corresponding red-detuned sideband at $\delta/(2\pi)\approx-40$~kHz is suppressed because the atoms have been prepared predominantly in the $n_z=0$ ground band. The inter-band transitions can also be used to selectively prepare the atoms in specific Bloch bands. For example, in order to prepare atoms exclusively in the $n_z=1$ band, a clock laser pulse is applied to the $\ket{g}_0\to\ket{e}_1$ blue sideband transition. A strong ``clearing" pulse is then applied to remove any remaining atoms in $\ket{g}$, leaving atoms in $\ket{e}$ unperturbed. The remaining atoms are thus purified in the $n_z=1$ Bloch band, and can be used for further experiments or measurements.

\begin{figure}
\begin{center}
\includegraphics[width=4.5in]{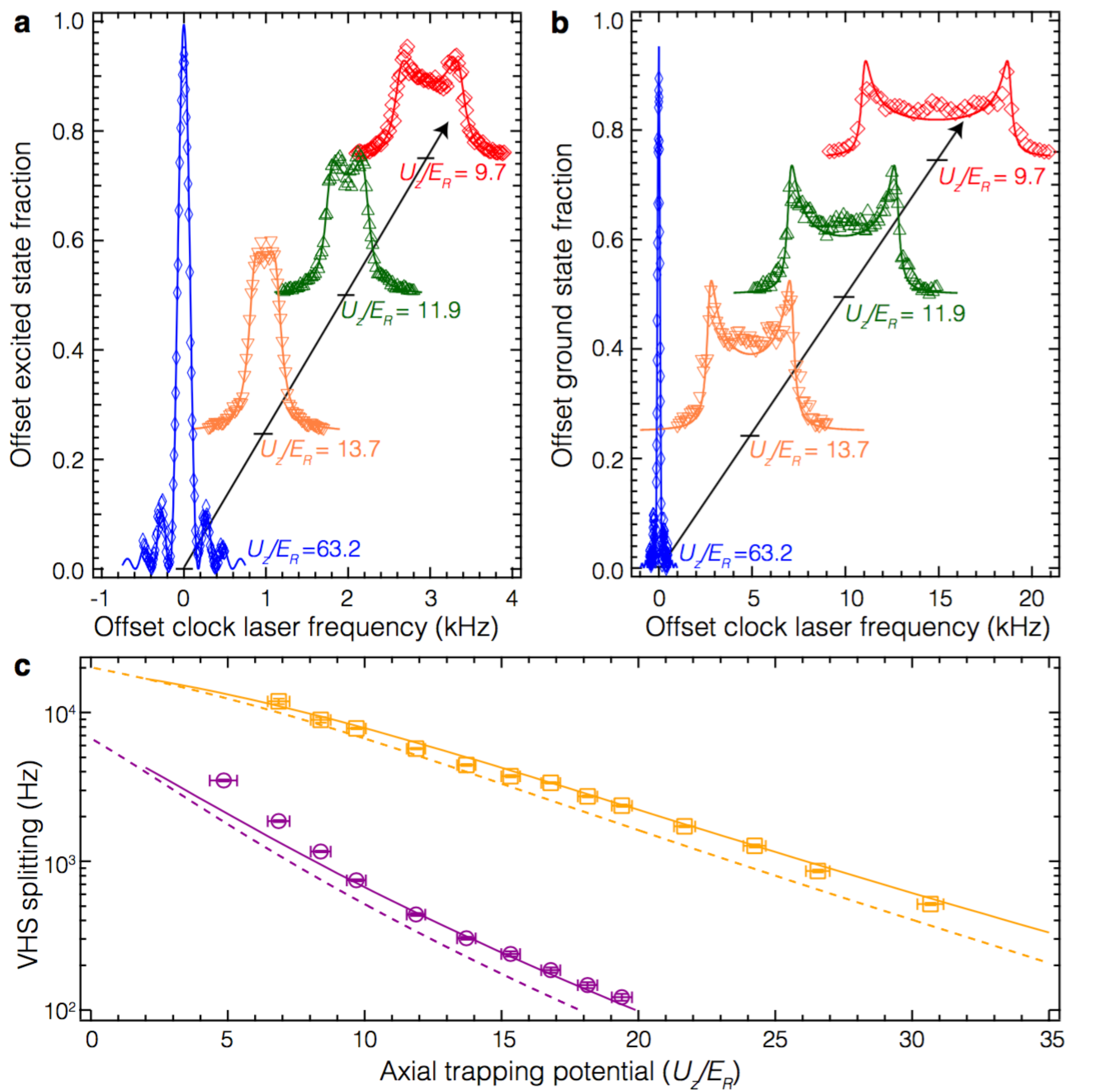}
\end{center}
 \caption{\textbf{Van Hove singularities and band mapping.} \textbf{a,} Excited state fraction following a $\pi$-pulse ($\Omega=2\pi\times100$~Hz) at four axial trapping potentials, with the atoms initially prepared in $\ket{g}_0$. The curves are offset in both $x$ and $y$ for clarity. The split peaks at $U_{z}/E_{r}\le13.7$ are a result of divergences in the atomic density of states known as Van Hove singularities (VHSs). \textbf{b,} Ground state fraction following a Rabi pulse for the same potentials shown in \textbf{a}, with the atoms initially prepared in $\ket{e}_1$. The duration of the Rabi pulse was varied to improve population contrast (see Methods). \textbf{c,} The splitting between the VHS peaks in the $\ket{g}_0\to\ket{e}_0$ (purple circles) and $\ket{g}_1\to\ket{e}_1$ (orange squares) transition lineshapes as a function of trapping potential, extracted using fits as shown in \textbf{a} and \textbf{b}. Horizontal error bars are 1-$\sigma$ standard error estimates from spectroscopy of the axial sidebands, vertical error bars are 1-$\sigma$ standard error estimates for the extracted VHS splitting. The no-free parameter predicted VHS splittings for atoms in the ground and first excited bands of a 1D sinusoidal lattice (purple and orange dashed lines respectively,) and for a model including the transverse motional modes and finite atomic temperatures (purple and orange solid lines) are shown for comparison.}
\label{fig_2}%
\end{figure}

In Fig.~\ref{fig_2}, we demonstrate the use of selective band preparation to probe the impact of SOC on the $\ket{g}_0\to\ket{e}_0$ and $\ket{e}_1\to\ket{g}_1$ carrier transitions. In Fig.~\ref{fig_2}a, the atoms are initially prepared in $\ket{g}_0$, and a $\pi$-pulse of the clock laser is applied. At $U_{z}/E_{r}=63.2$ (narrow blue diamonds) the result is a typical Fourier-limited Rabi lineshape of the $\ket{g}_0\to\ket{e}_0$ transition. However, as $U_z$ is reduced (in the regime $U_z>E_r$), the transition begins to broaden and split into two peaks, with the splitting scaling proportionally to the tunneling rate in the lowest Bloch band, which scales\cite{ManyBodyReview} as $J\approx (4/\hbar\sqrt{\pi})E_r(U_z/E_r)^{3/4}\exp[{-2\sqrt{U_z/E_r}}]$. As shown in Fig.~\ref{fig_2}b, the same behavior is observed when the atoms are initially prepared in the $\ket{e}_1$ state, with the $\ket{e}_1\to\ket{g}_1$ transition exhibiting much larger splittings compared to the $\ket{g}_0\to\ket{e}_0$ transition for the same axial potential.

The split lineshapes of the clock carrier transition at low $U_z$, which have been theoretically predicted\cite{WallSOC, Lemonde}, can be understood by considering the band dispersion curves presented in Fig.~\ref{fig_1}b. Because the $\ket{g}_0$ and $\ket{e}_0$ bands are shifted with respect to each other in quasimomentum by $\phi$, the transition frequency is $q$ dependent. In the tight-binding approximation, the largest momentum-induced detuning from the bare clock transition frequency is given by $\Delta=4J|\sin(\phi/2)|$, where $4J$ is the bandwidth of the $n_z=0$ ground band, and $\sin(\phi/2)=0.97\approx1$, resulting in an overall broadening of the transition by $2\Delta\approx8J$. The probability of a transition between the two bands at a specific $q$ is then determined by the joint transition density of states, which diverges at saddle points in the energy difference between the band dispersion curves. These divergence points in the density of states of a crystalline lattice are called Van Hove singularities (VHSs), and are well known from optical absorption spectra in solids and scanning tunneling microscopy\cite{vanHove,vanHove2}. The measured lineshapes are a convolution of the atomic transition density of states with the Rabi lineshape for a single atom (see Methods). In Fig.~\ref{fig_2}a and b, two VHS peaks are visible at $\delta=\pm\Delta$ when $\Delta>\Omega$, while at higher trapping potentials $\Delta\ll\Omega$ and the two VHS peaks merge into the standard Rabi lineshape.

Optical clock spectroscopy also provides a direct, \textit{in situ} probe of the Bloch bandwidths, and thus of the tunneling rate $J$, through the VHS splitting. We demonstrate the power of this technique by measuring Rabi lineshapes at a range of axial potentials from $U_{z}/E_{r}=5.5$ to $U_{z}/E_{r}=63.2$, and fitting a convolution of the joint transition density of states for the SOC bands with the measured Rabi lineshape at high $U_{z}/E_{r}$ to extract the VHS splitting (with example fits shown in Fig.~\ref{fig_2}a and b, see Methods). The extracted splittings for the $\ket{g}_0\to\ket{e}_0$ (purple circles) and $\ket{e}_1\to\ket{g}_1$ (orange squares) transitions are plotted as a function of trapping potential in Fig.~\ref{fig_2}c. Only splittings for which the two peaks were resolvable are shown. At $U_{z}/E_{r}=5.5$ the VHS splitting of the $\ket{e}_1\to\ket{g}_1$ transition overlapped with the inter-band transitions and the splitting could not be unambiguously extracted. The no-free parameter VHS splitting anticipated for atoms in a 1D lattice (purple and orange dashed lines respectively for $n_z=0$ and $n_z=1$), as well as the splitting predicted by the perturbative model (solid lines), are shown for comparison. The origin of the slight difference between the experimental data and the model prediction in the ground band at lower trapping potentials may be related to inhomogeneities in the axial trapping potential across the populated lattice sites, or an incomplete theory description of the transverse/longitudinal coupling.

\begin{figure}
\begin{center}
\includegraphics[width=4.5in]{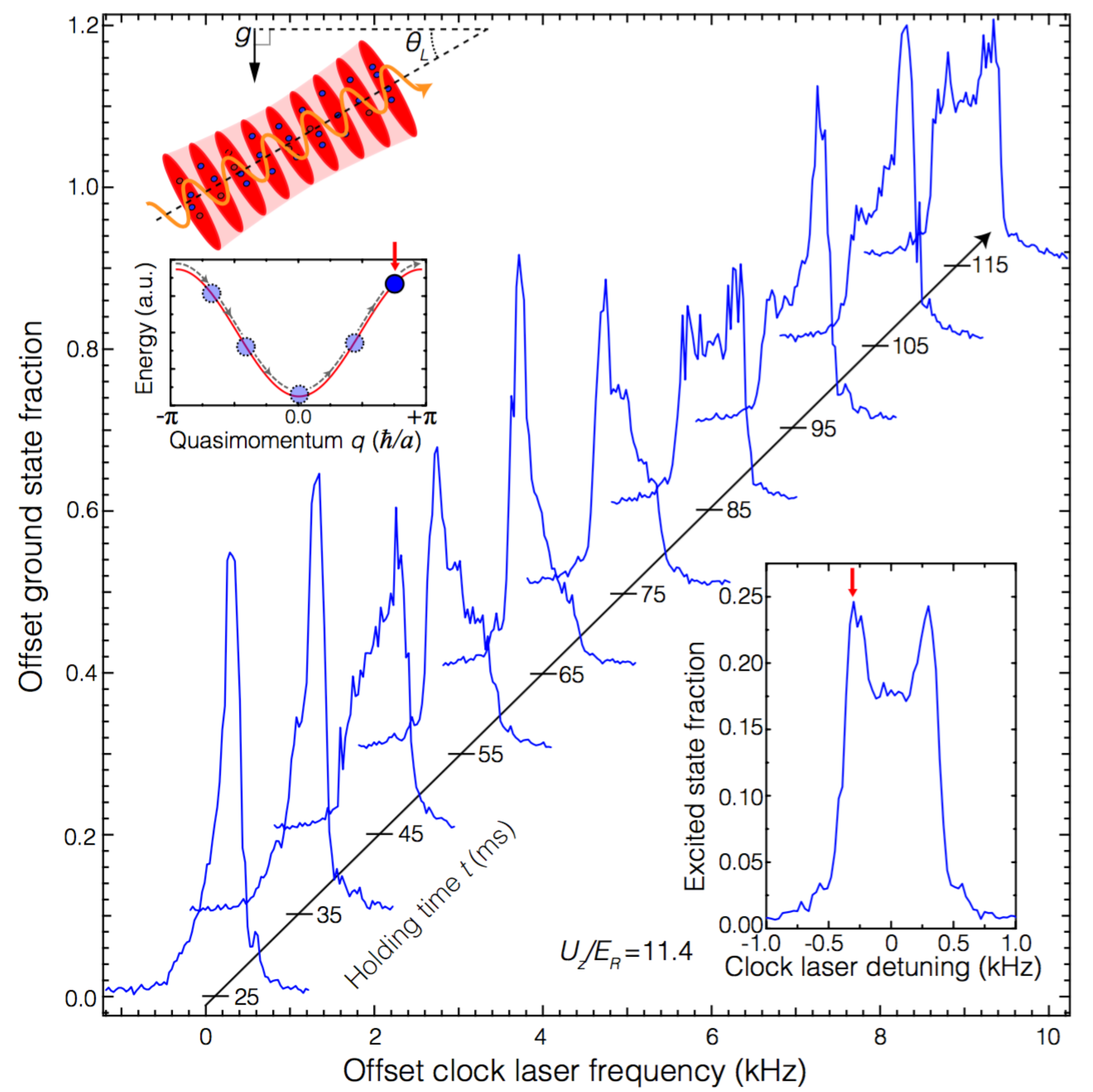}
\end{center}
 \caption{\textbf{Bloch oscillations.} \textbf{(Lower right inset,)} The split Rabi lines shown in Fig.~2a,b enable the spectral selection of atoms with a particular quasimomentum $q^{*}$. A $\pi$-pulse ($\Omega=2\pi\times100$~Hz) tuned to the left Van Hove peak is applied (red arrow) and the remaining ground state atoms are removed. \textbf{(Upper left inset,)} In a lattice tilted with respect to gravity, the atoms initially prepared with corresponding $q^{*}\approx\pi$ (red arrow) undergo Bloch oscillations. \textbf{(Main figure,)} Following a variable wait time $t$ from the end of the first $\pi$-pulse, a second $\pi$-pulse is applied, revealing a highly asymmetric lineshape (first blue curve at 25 ms). The atoms undergo Bloch oscillations, resulting in periodic oscillations of the lineshape as the waiting time between the first and second pulse is varied. The curves are offset in both $x$ and $y$ for clarity.}
\label{fig_3}%
\end{figure}

Just as the spectroscopically resolved sidebands enabled band preparation, the quasimomentum dependence of the clock transition frequency enables the selective preparation and subsequent manipulation of atoms with particular quasimomenta (Fig.~\ref{fig_3}). Following initialization in the $\ket{g}_0$ state, a clock pulse with Rabi frequency $\Omega<2\Delta$ is applied to the carrier transition with a detuning $|\delta^{*}|\le\Delta$. Only atoms with quasimomenta in a window centered around $q^{*}\approx\arccos{(\delta^{*}/\Delta})$ with a width bounded by $2\pi\Omega/\Delta$ will be excited to $\ket{e}_0$, while atoms with quasimomenta outside this window will be left in $\ket{g}$. A strong ``clearing" pulse is applied to remove atoms in $\ket{g}$, leaving only the atoms in $\ket{e}_0$ with quasimomenta in the window centered around $q^{*}$. Following a variable wait time $t$, a second $\pi$-pulse is used to measure the lineshape. If the lattice is tilted with respect to gravity, during the wait time the atoms will undergo Bloch oscillations\cite{BlochOsc1}, with their quasimomenta evolving as $q(t)=q_{o}+\nu_{B}t$, where $q_{o}$ is the initial quasimomentum of the atom, and the value of $q(t)$ is restricted to the first Brillouin zone. The Bloch oscillation frequency $\nu_{B}$ is given by $\nu_{B}=(mg\lambda_{L}\sin(\theta_{L}))/(4\pi\hbar)$, where $g$ is the acceleration due to gravity, and $\theta_{L}$ is the angle of tilt of the lattice. In this \textit{in situ} observation of Bloch oscillations in a tilted lattice\cite{GreinerBloch}, the highly asymmetric lineshapes oscillate back and forth as the quasimomenta cycles through the Brillouin zone at a frequency of $\nu_{B}=14$~Hz,  corresponding to a lattice tilt of $\theta_{L}=16$~milliradians.

\begin{figure}
\begin{center}
\includegraphics[width=4.5in]{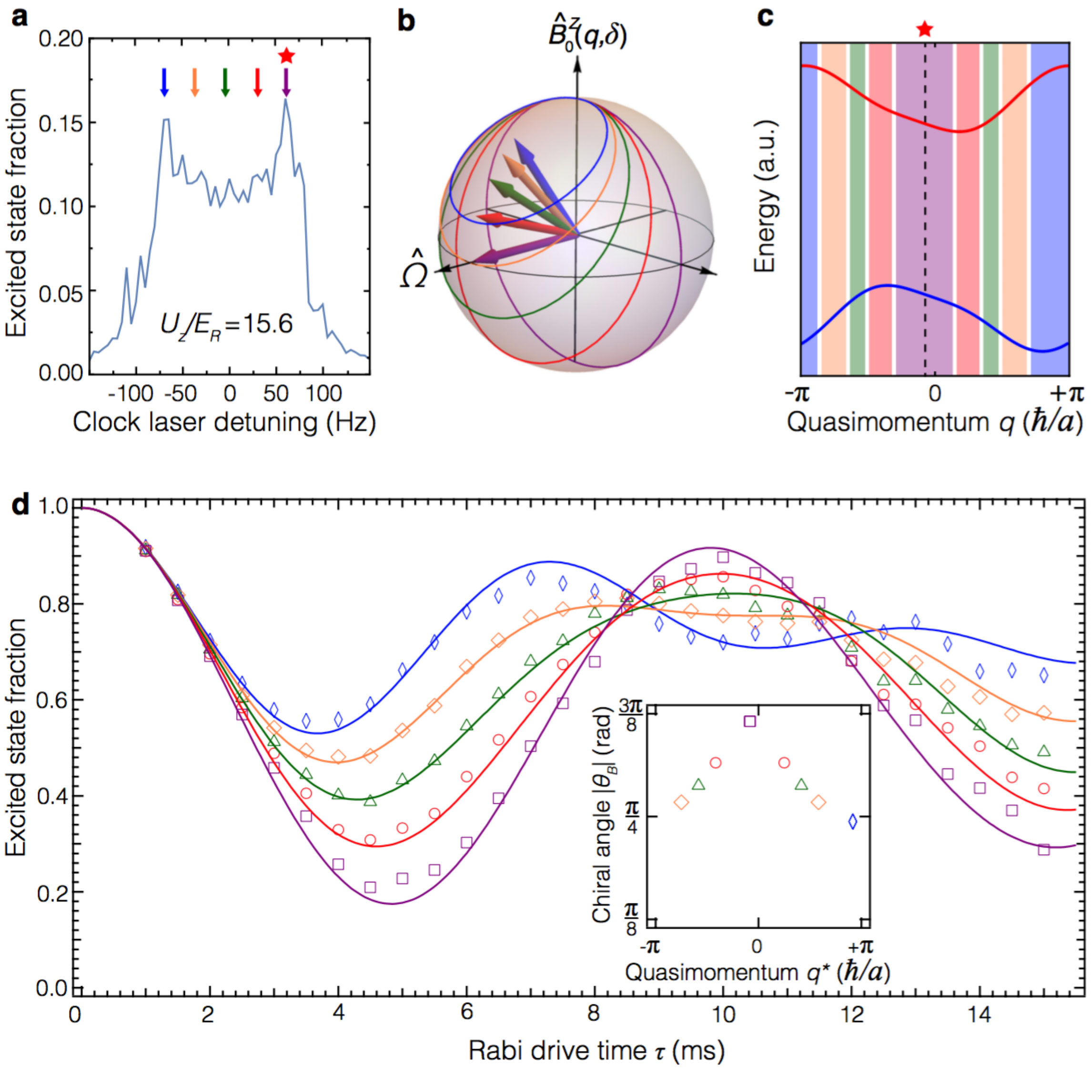}
\end{center}
 \caption{\textbf{Rabi measurements of the chiral Bloch vector.} \textbf{a,} Atoms are selectively prepared in a quasimomenta window using a $\pi$-pulse ($\Omega=2\pi\times10$~Hz,) with five different pulse detunings (colored arrows). An axial potential of $U_{z}/E_{r}=15.6$ is used, resulting in $\Delta/(2\pi)\approx67$~Hz. A second stronger Rabi pulse ($\Omega=2\pi\times100$~Hz) tuned to the right Van Hove peak is used to generate SOC (red star). \textbf{b,} The chiral Bloch vectors corresponding to the detunings in \textbf{a} are shown, along with the clock state spin precession for each vector. \textbf{c,} The SOC band structure experienced by the atoms during the second pulse, with each quasimomenta window color coded to match the detunings in \textbf{a}. The selection windows overlap, with the width of each window only intended to illustrate their relative values.  \textbf{d,} Excited state fraction as a function of duration of the second pulse for the five initial pulse detunings shown in \textbf{a} (data points), along with the no-free parameters dynamics predicted by a semi-classical model (colored solid lines, see Methods). \textbf{(Inset,)} Corresponding extracted chiral Bloch vector angle for each quasimomenta window centered at $q^{*}$.}
\label{fig_4}%
\end{figure}

We characterize the $q$ dependence of the chiral Bloch vector angle $\theta_B$ (Eq.~\ref{eq:theta_B}) using the same quasimomentum selection technique used to observe Bloch oscillations. For these measurements the lattice tilt was adjusted to minimize $\nu_{B}\le3$~Hz, with $\theta_{L}\le3.5$~milliradians.  As shown in Fig.~\ref{fig_4}a, atoms are prepared in $\ket{e}_0$ with quasimomenta $q^{*}$. In five separate experiments $q^{*}$ is varied using a range of detunings $\delta^{*}$ spanning the two VHS peaks (colored arrows). A strong Rabi pulse of duration $\tau$ is applied with detuning $\delta$ corresponding to the right VHS peak (red star), generating SOC with the corresponding $q^{*}$-dependent chiral Bloch vector (pointing along the direction of the effective magnetic field $\vec{B}_{n_z}(\Omega,q,\delta)$) shown in Fig.~\ref{fig_4}b, and with the SOC band structure shown in Fig.~\ref{fig_4}c. In Fig.~\ref{fig_4}d, the resulting population fraction in $\ket{e}$ for each prepared $q^{*}$ (with corresponding color coding) is plotted as a function of the evolution time $\tau$. The dynamics are entirely captured by the $q^{*}$-dependent spin precessions about the chiral Bloch vectors depicted in Fig.~\ref{fig_4}b. The theoretical calculations involve no free parameters and use only the experimental values of $\delta^{*}$, $\delta$, $\nu_{B}$, and $\Omega$ (colored solid lines in Fig.~\ref{fig_4}d, see Methods). The dephasing of the spin precession at longer $\tau$ is well described by the known initial distribution of quasimomenta $q^{*}$ (see Methods), and could be mitigated by reducing the Rabi frequency used for the $q^{*}$ selection with respect to the VHS splitting $2\Delta$, at the cost of reduced signal to noise ratio due to the smaller number of atoms selected. In the inset of Fig.~\ref{fig_4}d we plot the corresponding extracted chiral Bloch vector angle $|\theta_B|$ for each initial pulse detuning as a function of prepared quasimomenta $q^{*}$. Because there exists a one-to-one correspondence between the topological chiral edge modes of the two-dimensional Hofstadter model\cite{aidelsburger,ShakenLattice2} and the energy bands and eigenstates of the synthetic ladder\cite{ChiralLadders,BlochChiral} shown in Fig.~\ref{fig_1}a, the $q^{*}$ dependence of $|\theta_B|$ that we measure spectroscopically is a direct manifestation of the well-defined chirality of the edge states of the Hofstadter Hamiltonian. The presence of spin-momentum locking in the ladder eigenmodes has been previously observed using TOF\cite{BlochChiral,SpielmanSOC,FallaniSOC}.

In this work we have implemented and characterized SOC and a synthetic momentum-dependent magnetic field with fermions in an OLC. We observed clean experimental signatures of SOC physics at $\mu$K temperatures without cooling to Fermi degeneracy, and observed no decoherence or heating at timescales of hundreds of milliseconds. The recent realization of a Fermi-degenerate 3D OLC\cite{Campbell} opens a window to the implementation of two and three-dimensional SOC, tuning of the SOC phase $\phi$, and the lower temperatures required for the preparation of novel many-body states\cite{SpielmanReview,DalibardReview,galitskiSOCreview,zhaiSOCreview,KondoLattice}. While this work has focused entirely on single particle physics, many-body correlations and SU(N) symmetry have been previously observed in OLCs\cite{ManyBodySr,SUN}, offering exciting prospects for studying the interplay between SOC and interactions in higher synthetic dimensions\cite{SpielmanReview,DalibardReview,zhaiSOCreview,galitskiSOCreview,LewensteinReview,MagneticXtals}.\\

\spacing{1.0}

\bibliographystyle{naturemag}
\bibliography{SOC}

\begin{thebibliography}{10}
\expandafter\ifx\csname url\endcsname\relax
  \def\url#1{\texttt{#1}}\fi
\expandafter\ifx\csname urlprefix\endcsname\relax\def\urlprefix{URL }\fi
\providecommand{\bibinfo}[2]{#2}
\providecommand{\eprint}[2][]{\url{#2}}

\bibitem{DalibardReview}
\bibinfo{author}{Dalibard, J.}, \bibinfo{author}{Gerbier, F.},
  \bibinfo{author}{Juzeli{\=u}nas, G.} \& \bibinfo{author}{{\"O}hberg, P.}
\newblock \bibinfo{title}{Colloquium: Artificial gauge potentials for neutral
  atoms}.
\newblock \emph{\bibinfo{journal}{Reviews of Modern Physics}}
  \textbf{\bibinfo{volume}{83}}, \bibinfo{pages}{1523} (\bibinfo{year}{2011}).

\bibitem{SpielmanReview}
\bibinfo{author}{Goldman, N.}, \bibinfo{author}{Juzeli{\=u}nas, G.},
  \bibinfo{author}{{\"O}hberg, P.} \& \bibinfo{author}{Spielman, I.}
\newblock \bibinfo{title}{Light-induced gauge fields for ultracold atoms}.
\newblock \emph{\bibinfo{journal}{Reports on Progress in Physics}}
  \textbf{\bibinfo{volume}{77}}, \bibinfo{pages}{126401}
  (\bibinfo{year}{2014}).

\bibitem{ManyBodyReview}
\bibinfo{author}{Bloch, I.}, \bibinfo{author}{Dalibard, J.} \&
  \bibinfo{author}{Zwerger, W.}
\newblock \bibinfo{title}{Many-body physics with ultracold gases}.
\newblock \emph{\bibinfo{journal}{Reviews of Modern Physics}}
  \textbf{\bibinfo{volume}{80}}, \bibinfo{pages}{885} (\bibinfo{year}{2008}).

\bibitem{galitskiSOCreview}
\bibinfo{author}{Galitski, V.} \& \bibinfo{author}{Spielman, I.~B.}
\newblock \bibinfo{title}{Spin-orbit coupling in quantum gases}.
\newblock \emph{\bibinfo{journal}{Nature}} \textbf{\bibinfo{volume}{494}},
  \bibinfo{pages}{49--54} (\bibinfo{year}{2013}).

\bibitem{zhaiSOCreview}
\bibinfo{author}{Zhai, H.}
\newblock \bibinfo{title}{Degenerate quantum gases with spin--orbit coupling: a
  review}.
\newblock \emph{\bibinfo{journal}{Reports on Progress in Physics}}
  \textbf{\bibinfo{volume}{78}}, \bibinfo{pages}{026001}
  (\bibinfo{year}{2015}).

\bibitem{SpielmanSOC}
\bibinfo{author}{Stuhl, B.}, \bibinfo{author}{Lu, H.-I.},
  \bibinfo{author}{Aycock, L.}, \bibinfo{author}{Genkina, D.} \&
  \bibinfo{author}{Spielman, I.}
\newblock \bibinfo{title}{Visualizing edge states with an atomic {Bose} gas in
  the quantum {Hall} regime}.
\newblock \emph{\bibinfo{journal}{Science}} \textbf{\bibinfo{volume}{349}},
  \bibinfo{pages}{1514--1518} (\bibinfo{year}{2015}).

\bibitem{FallaniSOC}
\bibinfo{author}{Mancini, M.} \emph{et~al.}
\newblock \bibinfo{title}{Observation of chiral edge states with neutral
  fermions in synthetic {Hall} ribbons}.
\newblock \emph{\bibinfo{journal}{Science}} \textbf{\bibinfo{volume}{349}},
  \bibinfo{pages}{1510--1513} (\bibinfo{year}{2015}).

\bibitem{BlochChiral}
\bibinfo{author}{Atala, M.} \emph{et~al.}
\newblock \bibinfo{title}{Observation of chiral currents with ultracold atoms
  in bosonic ladders}.
\newblock \emph{\bibinfo{journal}{Nature Physics}}
  \textbf{\bibinfo{volume}{10}}, \bibinfo{pages}{588--593}
  (\bibinfo{year}{2014}).

\bibitem{LevSOC}
\bibinfo{author}{Burdick, N.~Q.}, \bibinfo{author}{Tang, Y.} \&
  \bibinfo{author}{Lev, B.~L.}
\newblock \bibinfo{title}{A long-lived spin-orbit-coupled degenerate dipolar
  {Fermi} gas}.
\newblock \emph{\bibinfo{journal}{arXiv preprint arXiv:1605.03211}}
  (\bibinfo{year}{2016}).

\bibitem{GyuBoong}
\bibinfo{author}{Song, B.} \emph{et~al.}
\newblock \bibinfo{title}{Spin-orbit coupled two-electron {Fermi} gases of
  ytterbium atoms}.
\newblock \emph{\bibinfo{journal}{arXiv preprint arXiv:1608.00478}}
  (\bibinfo{year}{2016}).

\bibitem{KetterleSpinTex}
\bibinfo{author}{Li, J.} \emph{et~al.}
\newblock \bibinfo{title}{Spin-orbit coupling and spin textures in optical
  superlattices}.
\newblock \emph{\bibinfo{journal}{arXiv preprint arXiv:1606.03514}}
  (\bibinfo{year}{2016}).

\bibitem{WallSOC}
\bibinfo{author}{Wall, M.~L.} \emph{et~al.}
\newblock \bibinfo{title}{Synthetic spin-orbit coupling in an optical lattice
  clock}.
\newblock \emph{\bibinfo{journal}{Physical Review Letters}}
  \textbf{\bibinfo{volume}{116}}, \bibinfo{pages}{035301}
  (\bibinfo{year}{2016}).

\bibitem{ChiralLadders}
\bibinfo{author}{H{\"u}gel, D.} \& \bibinfo{author}{Paredes, B.}
\newblock \bibinfo{title}{Chiral ladders and the edges of quantum {Hall}
  insulators}.
\newblock \emph{\bibinfo{journal}{Physical Review A}}
  \textbf{\bibinfo{volume}{89}}, \bibinfo{pages}{023619}
  (\bibinfo{year}{2014}).

\bibitem{aidelsburger}
\bibinfo{author}{Aidelsburger, M.} \emph{et~al.}
\newblock \bibinfo{title}{Realization of the {Hofstadter} {Hamiltonian} with
  ultracold atoms in optical lattices}.
\newblock \emph{\bibinfo{journal}{Physical review letters}}
  \textbf{\bibinfo{volume}{111}}, \bibinfo{pages}{185301}
  (\bibinfo{year}{2013}).

\bibitem{ShakenLattice2}
\bibinfo{author}{Miyake, H.}, \bibinfo{author}{Siviloglou, G.~A.},
  \bibinfo{author}{Kennedy, C.~J.}, \bibinfo{author}{Burton, W.~C.} \&
  \bibinfo{author}{Ketterle, W.}
\newblock \bibinfo{title}{Realizing the {Harper} {Hamiltonian} with
  laser-assisted tunneling in optical lattices}.
\newblock \emph{\bibinfo{journal}{Physical Review Letters}}
  \textbf{\bibinfo{volume}{111}}, \bibinfo{pages}{185302}
  (\bibinfo{year}{2013}).

\bibitem{LewensteinReview}
\bibinfo{author}{Celi, A.} \emph{et~al.}
\newblock \bibinfo{title}{Synthetic gauge fields in synthetic dimensions}.
\newblock \emph{\bibinfo{journal}{Physical Review Letters}}
  \textbf{\bibinfo{volume}{112}}, \bibinfo{pages}{043001}
  (\bibinfo{year}{2014}).

\bibitem{MagneticXtals}
\bibinfo{author}{Barbarino, S.}, \bibinfo{author}{Taddia, L.},
  \bibinfo{author}{Rossini, D.}, \bibinfo{author}{Mazza, L.} \&
  \bibinfo{author}{Fazio, R.}
\newblock \bibinfo{title}{Magnetic crystals and helical liquids in
  alkaline-earth fermionic gases}.
\newblock \emph{\bibinfo{journal}{Nature Communications}}
  \textbf{\bibinfo{volume}{6}}, \bibinfo{pages}{8134} (\bibinfo{year}{2015}).

\bibitem{ClockSr}
\bibinfo{author}{Bloom, B.~J.} \emph{et~al.}
\newblock \bibinfo{title}{An optical lattice clock with accuracy and stability
  at the 10$^{-18}$ level}.
\newblock \emph{\bibinfo{journal}{Nature}} \textbf{\bibinfo{volume}{506}},
  \bibinfo{pages}{71--75} (\bibinfo{year}{2014}).

\bibitem{ManyBodySr}
\bibinfo{author}{Martin, M.~J.} \emph{et~al.}
\newblock \bibinfo{title}{A quantum many-body spin system in an optical lattice
  clock}.
\newblock \emph{\bibinfo{journal}{Science}} \textbf{\bibinfo{volume}{341}},
  \bibinfo{pages}{632--636} (\bibinfo{year}{2013}).

\bibitem{Blatt}
\bibinfo{author}{Blatt, S.} \emph{et~al.}
\newblock \bibinfo{title}{{Rabi} spectroscopy and excitation inhomogeneity in a
  one-dimensional optical lattice clock}.
\newblock \emph{\bibinfo{journal}{Physical Review A}}
  \textbf{\bibinfo{volume}{80}}, \bibinfo{pages}{052703}
  (\bibinfo{year}{2009}).

\bibitem{Lemonde}
\bibinfo{author}{Lemonde, P.} \& \bibinfo{author}{Wolf, P.}
\newblock \bibinfo{title}{Optical lattice clock with atoms confined in a
  shallow trap}.
\newblock \emph{\bibinfo{journal}{Physical Review A}}
  \textbf{\bibinfo{volume}{72}}, \bibinfo{pages}{033409}
  (\bibinfo{year}{2005}).

\bibitem{vanHove}
\bibinfo{author}{Van~Hove, L.}
\newblock \bibinfo{title}{The occurrence of singularities in the elastic
  frequency distribution of a crystal}.
\newblock \emph{\bibinfo{journal}{Physical Review}}
  \textbf{\bibinfo{volume}{89}}, \bibinfo{pages}{1189} (\bibinfo{year}{1953}).

\bibitem{vanHove2}
\bibinfo{author}{Kim, P.}, \bibinfo{author}{Odom, T.~W.},
  \bibinfo{author}{Huang, J.-L.} \& \bibinfo{author}{Lieber, C.~M.}
\newblock \bibinfo{title}{Electronic density of states of atomically resolved
  single-walled carbon nanotubes: {Van} {Hove} singularities and end states}.
\newblock \emph{\bibinfo{journal}{Physical Review Letters}}
  \textbf{\bibinfo{volume}{82}}, \bibinfo{pages}{1225} (\bibinfo{year}{1999}).

\bibitem{BlochOsc1}
\bibinfo{author}{Dahan, M.~B.}, \bibinfo{author}{Peik, E.},
  \bibinfo{author}{Reichel, J.}, \bibinfo{author}{Castin, Y.} \&
  \bibinfo{author}{Salomon, C.}
\newblock \bibinfo{title}{{Bloch} oscillations of atoms in an optical
  potential}.
\newblock \emph{\bibinfo{journal}{Physical Review Letters}}
  \textbf{\bibinfo{volume}{76}}, \bibinfo{pages}{4508} (\bibinfo{year}{1996}).

\bibitem{GreinerBloch}
\bibinfo{author}{Preiss, P.~M.} \emph{et~al.}
\newblock \bibinfo{title}{Strongly correlated quantum walks in optical
  lattices}.
\newblock \emph{\bibinfo{journal}{Science}} \textbf{\bibinfo{volume}{347}},
  \bibinfo{pages}{1229--1233} (\bibinfo{year}{2015}).

\bibitem{Campbell}
\bibinfo{author}{Campbell, S.~L.} \emph{et~al.}
\newblock \bibinfo{title}{A {Fermi}-degenerate {3D} optical lattice clock}
  (\bibinfo{year}{2016}).
\newblock \bibinfo{note}{In preparation}.

\bibitem{KondoLattice}
\bibinfo{author}{Isaev, L.}, \bibinfo{author}{Schachenmayer, J.} \&
  \bibinfo{author}{Rey, A.}
\newblock \bibinfo{title}{Spin-orbit-coupled correlated metal phase in {Kondo}
  lattices: An implementation with alkaline-earth atoms}.
\newblock \emph{\bibinfo{journal}{Physical Review Letters}}
  \textbf{\bibinfo{volume}{117}}, \bibinfo{pages}{135302}
  (\bibinfo{year}{2016}).

\bibitem{SUN}
\bibinfo{author}{Zhang, X.} \emph{et~al.}
\newblock \bibinfo{title}{Spectroscopic observation of {SU(N)}-symmetric
  interactions in {Sr} orbital magnetism}.
\newblock \emph{\bibinfo{journal}{Science}} \textbf{\bibinfo{volume}{345}},
  \bibinfo{pages}{1467--1473} (\bibinfo{year}{2014}).

\bibitem{FallaniNew}
\bibinfo{author}{Livi, L.} \emph{et~al.}
\newblock \bibinfo{title}{Synthetic dimensions and spin-orbit coupling with an
  optical clock transition}.
\newblock \emph{\bibinfo{journal}{arXiv preprint arXiv:1609.04800}}
  (\bibinfo{year}{2016}).

\bibitem{GadwayNew}
\bibinfo{author}{An, F.~A.}, \bibinfo{author}{Meier, E.~J.} \&
  \bibinfo{author}{Gadway, B.}
\newblock \bibinfo{title}{Direct observation of chiral currents and magnetic
  reflection in atomic flux lattices}.
\newblock \emph{\bibinfo{journal}{arXiv preprint arXiv:1609.09467}}
  (\bibinfo{year}{2016}).

\end{thebibliography}


\begin{thebibliography}{1}
\setcounter{enumiv}{30}
\expandafter\ifx\csname url\endcsname\relax
  \def\url#1{\texttt{#1}}\fi
\expandafter\ifx\csname urlprefix\endcsname\relax\def\urlprefix{URL }\fi
\providecommand{\bibinfo}[2]{#2}
\providecommand{\eprint}[2][]{\url{#2}}

\bibitem{Arimondo}
\bibinfo{author}{Lignier, H.} \emph{et~al.}
\newblock \bibinfo{title}{Dynamical control of matter-wave tunneling in
  periodic potentials}.
\newblock \emph{\bibinfo{journal}{Physical Review Letters}}
  \textbf{\bibinfo{volume}{99}}, \bibinfo{pages}{220403}
  (\bibinfo{year}{2007}).

\bibitem{ShakenLattice3}
\bibinfo{author}{Parker, C.~V.}, \bibinfo{author}{Ha, L.-C.} \&
  \bibinfo{author}{Chin, C.}
\newblock \bibinfo{title}{Direct observation of effective ferromagnetic domains
  of cold atoms in a shaken optical lattice}.
\newblock \emph{\bibinfo{journal}{Nature Physics}}
  \textbf{\bibinfo{volume}{9}}, \bibinfo{pages}{769--774}
  (\bibinfo{year}{2013}).

\bibitem{Haldane}
\bibinfo{author}{Jotzu, G.} \emph{et~al.}
\newblock \bibinfo{title}{Experimental realization of the topological {Haldane}
  model with ultracold fermions}.
\newblock \emph{\bibinfo{journal}{Nature}} \textbf{\bibinfo{volume}{515}},
  \bibinfo{pages}{237--240} (\bibinfo{year}{2014}).

\bibitem{BerryCurve}
\bibinfo{author}{Fl{\"a}schner, N.} \emph{et~al.}
\newblock \bibinfo{title}{Experimental reconstruction of the {Berry} curvature
  in a {Floquet} {Bloch} band}.
\newblock \emph{\bibinfo{journal}{Science}} \textbf{\bibinfo{volume}{352}},
  \bibinfo{pages}{1091--1094} (\bibinfo{year}{2016}).

\bibitem{BandRelax1}
\bibinfo{author}{M{\"u}ller, T.}, \bibinfo{author}{F{\"o}lling, S.},
  \bibinfo{author}{Widera, A.} \& \bibinfo{author}{Bloch, I.}
\newblock \bibinfo{title}{State preparation and dynamics of ultracold atoms in
  higher lattice orbitals}.
\newblock \emph{\bibinfo{journal}{Physical Review Letters}}
  \textbf{\bibinfo{volume}{99}}, \bibinfo{pages}{200405}
  (\bibinfo{year}{2007}).

\bibitem{BandRelax2}
\bibinfo{author}{Spielman, I.~B.} \emph{et~al.}
\newblock \bibinfo{title}{Collisional deexcitation in a quasi-two-dimensional
  degenerate bosonic gas}.
\newblock \emph{\bibinfo{journal}{Physical Review A}}
  \textbf{\bibinfo{volume}{73}}, \bibinfo{pages}{020702}
  (\bibinfo{year}{2006}).

\bibitem{BishofInelastic}
\bibinfo{author}{Bishof, M.} \emph{et~al.}
\newblock \bibinfo{title}{Inelastic collisions and density-dependent excitation
  suppression in a ${}^{87}${Sr} optical lattice clock}.
\newblock \emph{\bibinfo{journal}{Physical Review A}}
  \textbf{\bibinfo{volume}{84}}, \bibinfo{pages}{052716}
  (\bibinfo{year}{2011}).

\bibitem{MartinThesis}
\bibinfo{author}{Martin, M.~J.}
\newblock \emph{\bibinfo{title}{Quantum metrology and many-body physics:
  pushing the frontier of the optical lattice clock}}.
\newblock Ph.D. thesis, \bibinfo{school}{CU Boulder} (\bibinfo{year}{2013}).

\bibitem{WallHazzardRey}
\bibinfo{author}{Wall, M.~L.}, \bibinfo{author}{Hazzard, K. R.~A.} \&
  \bibinfo{author}{Rey, A.~M.}
\newblock \bibinfo{title}{Effective many-body parameters for atoms in
  nonseperable {Gaussian} optical potentials}.
\newblock \emph{\bibinfo{journal}{Physical Review A}}
  \textbf{\bibinfo{volume}{92}}, \bibinfo{pages}{013610}
  (\bibinfo{year}{2015}).

\end{thebibliography}

\noindent\textbf{Acknowledgements}\\
We are grateful to N.R.~Cooper for insights and discussions, and S.L.~Campbell, N.~Darkwah Oppong, A.~Goban, R.B.~Hutson, D.X.~Reed, J.~Robinson, L.~Sonderhouse, and W.~Zhang for technical contributions and discussions. This research is supported by NIST, the NSF Physics Frontier Center at JILA (NSF-PFC-1125844), AFOSR-MURI, AFOSR, DARPA, and ARO. S.K., M.L.W., and G.E.M. acknowledge the NRC postdoctoral fellowship program.

\noindent\textbf{Authors' note:} Following the completion of this work, implementations of SOC using the direct optical transition in $^{173}$Yb atoms\cite{FallaniNew}, and a two-dimensional manifold of discrete atomic momentum states in $^{87}$Rb atoms\cite{GadwayNew} were reported. 

\noindent\textbf{Author contributions}\\
S.K., S.L.B., T.B., G.E.M., X.Z., and J.Y. contributed to the executions of the experiments. M.L.W., A.P.K., and A.M.R. contributed to the development of the theory model. All authors discussed the results, contributed to the data analysis and worked together on the manuscript.

\noindent\textbf{Author information}\\
The authors declare no competing financial interests. Correspondence and requests for materials should be addressed to S.K. (shimonk@jila.colorado.edu).


\spacing{1.5}

\end{document}


\Large
\noindent\textbf{Methods}\\
\normalsize
\noindent\textbf{Experimental methods}\\
\noindent\textbf{$\bullet$ Measurement protocol}\\
To load the optical lattice, $^{87}$Sr atoms are cooled with two sequential 3D magneto-optical-traps (MOTs), the first using the strong $^{1}$S$_0\to^{1}$P$_1$ transition (461 nm), and the second using the narrower $^{1}$S$_0\to^{3}$P$_1$ transition (689 nm). Following the second MOT, the atoms are sufficiently cold and dense to be trapped in the optical lattice. Once in the lattice, the atoms are first nuclear spin-polarized, and then further cooled using axial sideband and radial doppler cooling on the 689 nm transition.

At the end of each experiment the population of the $\ket{g}$ state is measured by driving the atoms on the 461 nm cycling transition and counting the emitted photons. After 5 ms all $\ket{g}$ atoms have been heated out of the lattice, at which point the atoms in $\ket{e}$ are optically pumped into the $\ket{g}$ state and are counted in the same manner. A 5 ms long 461 nm pulse is also used as the ``clearing" pulse to remove ground state atoms in the protocols used for preparing atoms in specific bands and with select quasimomenta.

\noindent\textbf{$\bullet$ Characterization of the trapping potential}\\
The cylindrically symmetric trapping potential experienced by an atom at position $z$ along the axis of propagation of the lattice beams and a distance $r$ from the center of the beams is given by
\begin{align}
\label{eq:FullPot}V\left(r,z\right)&=-\left(V_{\mathrm{const}}+U_z\cos^2\left(k_L z\right)\right)e^{-2 r^2/w_0^2}\, ,
\end{align}
\noindent where $V_{\mathrm{const}}={\alpha\left(\lambda_{L}\right)}{}\left(P_{1}+P_{2}-2\sqrt{P_{1}P_{2}}\right)/(\pi \varepsilon_0 c w_0^2)$, $U_z=4 \alpha\left(\lambda_{L}\right)\sqrt{P_{1}P_{2}}/(\pi \varepsilon_0 c w_0^2)$, $k_L=2\pi/\lambda_L$, $\varepsilon_0$ is the permittivity of free space, $c$ the speed of light, and $\alpha\left(\lambda_{L}\right)$ is the AC polarizability evaluated at $\lambda_{L}$. Because $P_{1} \gg P_{2}$, to first order the trapping potential in the radial direction is proportional to $V_{\mathrm{const}}\propto P_{1},$  while the periodic axial trapping potential $U_{z}\propto \sqrt{P_{1}P_{2}}$. In contrast to prior experiments that generate synthetic gauge fields by periodically shaking the optical lattice potential\cite{Arimondo,ShakenLattice3,Haldane,BerryCurve}, here the lattice potential is kept constant, and it is the probing laser itself that induces the SOC.

\begin{figure}
 \centering
\includegraphics[width=0.99\columnwidth]{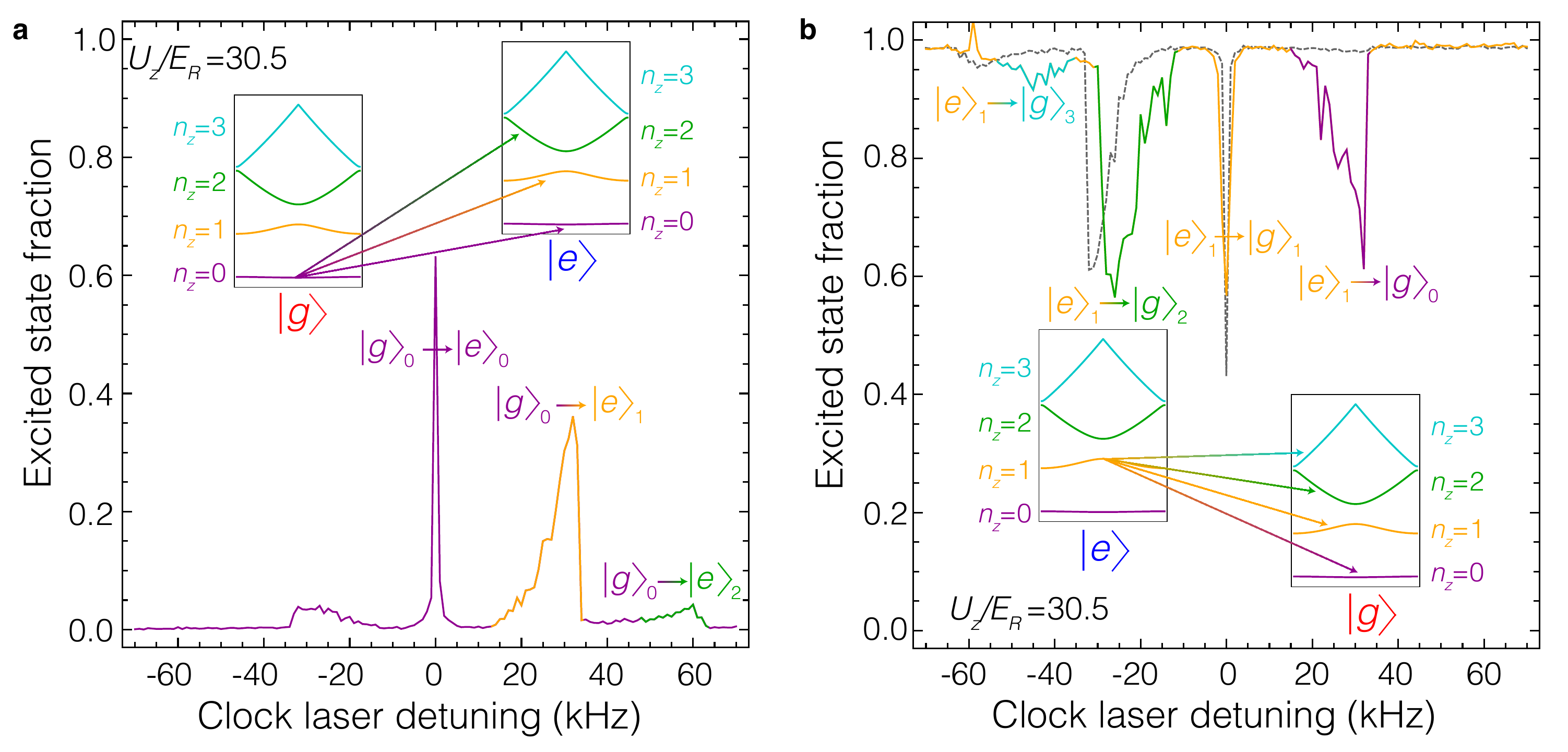}
\caption{\textbf{a,} Spectroscopy of atoms prepared in $\ket{g}_0$ reveals the band spacing of the lattice, with inter-band transitions color-coded by the final band. \textbf{b,} Spectroscopy of atoms in $\ket{e}_1$, prepared by driving the $\ket{g}_0\to\ket{e}_1$ transition shown in \textbf{a}, then removing any remaining atoms in $\ket{g}$. The spectrum for atoms in $\ket{e}_0$ from Fig.~1d in the main text is shown for comparison (dashed gray line).}\label{fig:ED1}
\end{figure}

\noindent\textbf{$\bullet$ Band preparation}\\
In Fig.~ED\ref{fig:ED1} (Extended Data), we demonstrate how the inter-band transitions can be used for band preparation. Fig.~ED\ref{fig:ED1}a shows spectroscopy of the carrier and inter-band transitions from the initial $\ket{g}_0$ state. In order to prepare atoms exclusively in the $n_z=1$ band, a clock laser pulse is applied to the $\ket{g}_0\to\ket{e}_1$ blue sideband transition, which is at a detuning $\delta/2\pi\approx\nu_z-E_{r}/2\pi\hbar=35$~kHz for $U_{z}/E_{r}=30.5$. A strong ``clearing" pulse is then applied to remove any remaining atoms in $\ket{g}$, leaving atoms in $\ket{e}$ unperturbed. The remaining atoms are thus purified in the $n_z=1$ Bloch band, and can be used for further experiments or measurements. Fig.~ED\ref{fig:ED1}b shows spectroscopy of the sidebands following this protocol, with the anharmonicity of the band spacing resulting in an unequal frequency spacing between the $\ket{e}_1\to\ket{g}_2$ and $\ket{e}_1\to\ket{g}_0$ sideband transitions about the $\ket{e}_1\to\ket{g}_1$ carrier transition. The transitions out of $\ket{e}_0$ from Fig. 1d in the main text are shown for comparison. In contrast to prior experiments\cite{BandRelax1,BandRelax2} no band relaxation has been observed out of the $\ket{e}_1$ state over a waiting time greater than $500$ ms, due to the dilute atomic conditions achieved in the current experiment.

\noindent\textbf{$\bullet$ Measurements of axial heating and loss rates}\\
The axial heating rate in our lattice was measured using spectroscopy of the axial motional sidebands. The atoms were prepared in $\ket{e}_0$, and the clock laser was applied on resonance with the carrier transition for a variable wait time of up to 155 ms. The atoms in $\ket{g}$ were removed with a clearing pulse, and spectroscopy of the remaining atoms in $\ket{e}$ was performed. The axial sideband asymmetry was then used to determine the temperature and heating rate. The measurement was performed in both the strong confinement $(U_{z}/E_{r}\approx200)$ and tunneling allowed $(U_{z}/E_{r}\approx10)$ regimes. In both cases, the results were consistent with no axial heating over the 155 ms wait time. This is consistent with previous measurements of the temperature dependence of the clock transition coherence$^{20}$. 

The loss rates out of $\ket{e}$ were measured by preparing the atoms in $\ket{e}_0$, leaving the atoms in the dark for a variable wait time of up to 1.5~s, and then counting the number of atoms in the $\ket{e}$ and $\ket{g}$ states. Loss of atoms due to inelastic p-wave $e-e$ scattering was observed with a density dependent loss rate consistent with previous measurements\cite{BishofInelastic}. For the atomic densities used in this work, the measured loss rate corresponded to $\sim1.5$~s$^{-1}$, and thus did not have an impact on the  measurements presented here due to their shorter timescales. For future many-body experiments with higher desired densities, the temperature can be lowered by loading the lattice from a Fermi-degenerate gas$^{26}$, and the inelastic p-wave $e-e$ collisions will be significantly suppressed. An additional one body loss rate of $\sim0.2$~s$^{-1}$ was observed, consistent with the vacuum limited lifetime in the chamber. Finally, a $\ket{e}\to\ket{g}$ spin flip rate of $\sim0.1$~s$^{-1}$ was observed, consistent with previously measured two-photon Raman scattering from the lattice light via the ${}^{3}P_{1}$ state\cite{MartinThesis}. While this spin-flip rate is entirely negligible for the measurements presented here, it may set a limit on future SOC experiments, potentially requiring a different lattice geometry.

\noindent\textbf{Theoretical methods}\\
\noindent\textbf{$\bullet$ Perturbative model}\\
For the current experimental temperatures and loading conditions, to an excellent approximation, we can treat the coupling between the axial and radial degrees of freedom perturbatively and expand Eq.~\eqref{eq:FullPot} up to second order in $r$:
\begin{eqnarray}
V\left(r,z\right)&=& V_z(z)+ V_r(r)+ \eta V_{\mathrm{coupl}}(r,z)+\mathcal{O}\left(r^4\right)\end{eqnarray}
with $ V_z(z)=-U_z\cos^2\left(k_L z\right)$, $V_r(r)=m\omega_r^2 r^2/2 $, and $ V_{\mathrm{coupl}}(r,z)=- V_r(r) \sin^2\left(k_L z\right)$.  Here, $\omega_r=2\sqrt{\frac{U_z+V_{\mathrm{const}}}{m w_0^2}}$, $m$ is the atom mass,  and   $\eta=\frac{U_z}{U_z+V_{\mathrm{const}}}$ is an expansion parameter. To zeroth order in $\eta$  the Hamiltonian  is separable in the  $r$ and $z$ coordinates. In this limit the radial eigenfunctions  are harmonic oscillator functions, $ \langle {\bf r}| n_r, \upsilon\rangle=\phi_{n_r,\upsilon}\left({\bf r}\right)$ with eigenenergies $E_{n_r,\upsilon}=\hbar \omega_r \left(|\upsilon|+2n_r+1\right)$
 parameterized by   the radial quantum number $n_r=0,1,\dots$, and the  azimuthal quantum number $-n_r\leq \upsilon\leq n_r$.  The axial eigenfunctions are Bloch functions, $\langle z| n_z, q\rangle=\psi_{n_z,q}\left(z\right)$ with a band structure $E_{n_z}(q)$.
 For the cosinusoidal  potential in consideration  the latter  can be obtained analytically in terms of the even and odd Mathieu functions,  $\psi_{n_z,q}={\rm MathieuC}[a(q,U_z/(4E_r)), U_z/(4E_r), z ]+{\rm{i}} {\rm MathieuS}[a(q,U_z/(4E_r)), U_z/(4E_r), z ]$  and  $E_{n_z}(q)/E_r=a(q,U_z/(4E_r))$, where  the parameter $a$ is the characteristic Mathieu value  and for suitably chosen $n_z$-dependent ranges of $q$. 

Treating $V_{\mathrm{coupl}}(r,z)$ using first order perturbation theory in $\eta$ yields the energies:
\begin{equation}
\label{eq:perten}E_{n_z, n_r,\upsilon}(q)=E_{n_r,\upsilon}+E_{n_z}(q)-\frac{1}{2}\eta E_{n_r,\upsilon} \langle n_z,q|\sin^2\left(k_L z\right)| n_z, q\rangle .
\end{equation} This expression generalizes the one obtained by Blatt \emph{et al.}$^{20}$ in the deep lattice limit to include quasimomentum dependence of the energies and beyond-lowest-order anharmonic effects in the axial dimension.   We note that one can also analytically compute the first order corrections in terms of the unperturbed band energies using  the Feynman-Hellman theorem, $\langle n_z,q|\sin^2\left(k_L z\right)| n_z, q\rangle=\left(\frac{1}{2}+\frac{\partial E_{n_z}(q)}{\partial U_z}\right)$. This term has both $q-$independent and $q-$dependent components.  The former lead to thermally-dependent shifts of the mean band energies, and the latter gives a renormalization of the tunneling rate which depends on the radial temperature.  We have explicitly checked the validity of the perturbative energy expression by numerically finding the exact eigenstates of Eq.~\eqref{eq:FullPot}~\cite{WallHazzardRey}.

\noindent\textbf{$\bullet$ Line shapes}\\
To evaluate the lineshapes we perform a  thermal average using Boltzmann distributions with radial and axial temperatures $T_r$ and $T_z$, respectively.  The contribution to the lineshape from the $\ell^{\mathrm{th}}$ axial sideband, assuming atoms are initially populating  the $\ell_0$ band and internal state $\alpha=\pm$ (here $+$ is for  $g$ and $-$ is for $e$), is:
\begin{equation}
 P_{ \ell}^\pm=\sum_{n_z, n_r, \upsilon,q} \frac{q_{z}(\ell_0,n_z,q)q_{r}(\ell_0,n_r, \upsilon)}{Z_r(\ell_0)Z_z(\ell_0)}\left|\frac{\Omega_{ n_z,\ell,q}}{\Omega^{\pm \rm eff}_{n_z,\ell ,q,n_r,\upsilon}}\right|^2 \sin^2\left[\frac{t}{2}\Omega^{\pm \rm eff}_{n_z,\ell ,q,n_r,\upsilon}\right]
\label{eq:Rabi}
\end{equation}
\noindent where $\Omega^{\pm \rm eff}_{n_z,\ell,q,n_r,\upsilon}\equiv\sqrt{\left|\Omega_{n_z,\ell,q}\right|^2+\left(\pm \delta-(E_{n_z+\ell,n_r,\upsilon}(q+\phi)-E_{n_z,n_r,\upsilon}(q))/\hbar\right)^2}$ is the effective Rabi frequency,
$\Omega_{n_z,\ell ,q}\approx \Omega \langle n_z+\ell, q +\phi |e^{i 2\pi z/\lambda_c }|n_z,q\rangle$ the Rabi frequencies and  $\Omega$  the ``bare'' Rabi frequency. Note that in the regime where
the tight binding approximation is valid (i.e. not very shallow lattices) $\Omega_{n_z,\ell ,q}$ is a slowly varying function of $q$. On the other hand, it has a strong dependence on $\ell$ and for $|\ell|>0$ it is suppressed by the Lamb-Dicke parameter $\eta_{LD}=k_L \sqrt{\frac{ \hbar}{2 \omega_z m}}$, resulting in an effective Rabi frequency for the first sidebands, $\ell=\pm 1$, that is approximately an order of magnitude below that of the carrier.

We have also introduced the Boltzmann factors 
$q_z(\ell_0,n_z,q)\equiv \exp\left[-\beta_z\left(E_{n_z+\ell_0,0,0}(q)-E_{\ell_0,0,0}(0)\right)\right]$ and
$q_r(\ell_0,n_r, \upsilon)\equiv \exp\left[-\beta_r\sum_{q'}\left(E_{n_z+\ell_0,n_r,\upsilon}(q')-E_{n_z+\ell_0,0,0}(q')\right)/L\right]$  with
$\beta_{z,r}=1/k_B T_{z,r}$ and $L$ the total number of lattice sites. $Z_r(\ell_0)= \sum_{n_z, q} q_{z}(\ell_0,n_z,q)$ and
 and $Z_r(\ell_0)=\sum_{ n_r, \upsilon} q_{r}(\ell_0,n_r, \upsilon)$ are the radial and axial partition functions.
 
There are three leading  mechanisms that lead to broadening of the lineshape. One arises from the thermal population of different quasimomentum states combined with the finite momentum transfer by the probe laser when interrogating the clock states. This type of motion-induced broadening, which is a direct signature of the spin-orbit coupling mechanism, is the lattice analogue of Doppler broadening$^{21}$, generally discussed in  the spectra of free particles.
For the lattice case, instead of plane waves carrying specific  momentum, one needs to think about Bloch waves described  by two quantum numbers, the quasi-momentum and band index. The second broadnening mechanism is power broadening arising for strong Rabi pulses. For our spectroscopic parameters only the carrier transition is affected by it. The last source of broadening is the coupling between the radial and axial degrees of freedom, and thus is strongly  determined by the trapping frequencies and radial temperature. At short probing times (less than $\pi$ pulses), this type of broadening in the carrier transition  can be characterized by using a temperature dependent tunneling rate as we will explain below. For the sidebands, axial-radial coupling is the leading broadening mechanism$^{20}$ and it has been shown that it is well captured by the perturbative approach.

However, the  first-order perturbative approach neglects the radial dependence of the Rabi frequencies and radial sideband transitions induced by the laser. As discussed in detail in Blatt \emph{et al.}$^{20}$, the omitted higher order  terms can lead to a dephasing of the coherent oscillations in $P_{\ell}^\pm$ at times long compared to a $\pi$-pulse and induce additional broadening of the sideband spectra.  Nevertheless, those effects  can be accounted for   by using an effective radial temperature, which we  used as an effective fitting parameter, and by performing for the case of long probe times a time-average of Eq.~\eqref{eq:Rabi}, which yields
\begin{equation}
\bar{P}_{ \ell}^\pm= \frac{1}{2}\sum_{n_z, n_r, \upsilon,q} \frac{q_{z}(\ell_0,n_z,q)q_{r}(\ell_0,n_r, \upsilon)}{Z_r(\ell_0)Z_z(\ell_0)}\left|\frac{\Omega_{ n_z,\ell,q}}{\Omega^{\pm \rm eff}_{n_z,\ell ,q,n_r,\upsilon}}\right|^2. \label{ave}
\end{equation} The clock spectroscopy shown in Fig. 1c was taken by applying   the clock laser at high power ($\Omega=2\pi \times 200$ Hz)  and for many Rabi periods (during 80 ms) and thus we used Eq.~\ref{ave} to model the experiment and treated the   radial temperature as a fitting parameter. The radial  temperatures that provided the best fit  varied between $7-9$~$\mu$K and thus were higher than the experimentally measured radial temperatures through motional spectroscopy of the sidebands, consistent with previous detailed studies of the sideband lineshapes$^{20}$.

\begin{figure}
\centering
\includegraphics[width=0.99\columnwidth]{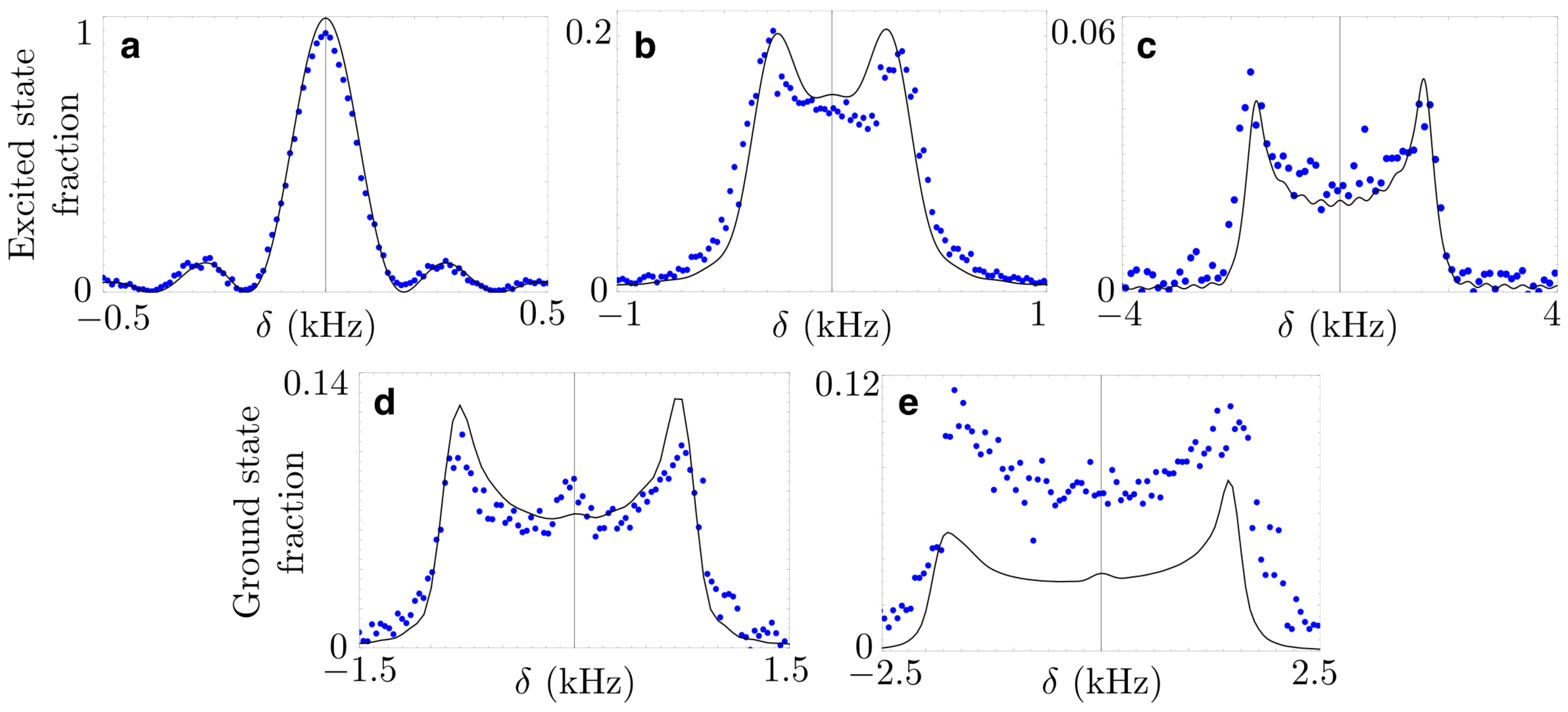}
\caption{Theoretical modeling of carrier lineshapes starting from the $|g\rangle_{0}$ (\textbf{a,b,c}) or the $|e\rangle_{1}$ states (\textbf{d,e}). For the ground band transitions  (\textbf{a,b,c}), all data were taken employing a $\pi$-pulse, and are well reproduced by a perturbative model. Three specific cases are shown covering from the deep lattice $U_z\sim 65~E_R$ regime (\textbf{a}), to the moderate lattice $U_z\sim10~E_R$ regime (\textbf{b}), all the way down to the shallow lattice $U_z\sim 3~E_R$ limit (\textbf{c}). For the first excited band transitions, the data in the deeper $U_z\sim 21.3~E_R$ lattice case, taken with a $\pi$-pulse, are well reproduced by the perturbative model (\textbf{d}). However, for the case of $U_z\sim 16~E_R$ where a longer pulse was employed, the perturbative theory, which ignores radial sideband transitions induced by the laser, only captures the width of the lineshape and not its amplitude (\textbf{e}).}\label{fig:Widths}
\end{figure}

For the carrier transition the dominant decoherence mechanism  arising from radial sideband transitions  manifests at long probing times as dephasing.  This effect is shown in Fig.~ED\ref{fig:Widths}, where it can be seen that the perturbative model, Eq.~\ref{eq:Rabi} with $l=0$, does an excellent job for pulses shorter than a $\pi$ pulse. However, for the longer pulse  employed (panel e), the perturbative theory captures only the width of the lineshape and not its amplitude.

\noindent\textbf{$\bullet$ Theoretical fit function to extract the VHS splittings:}  \\The coupling between the radial and axial degrees of freedom can be accounted for in the carrier lineshape by defining an effective thermally averaged tunneling rate. In the regime where the tight binding approximation is valid, $\langle \Delta E_{n_z,n_r,\upsilon}(q)\rangle\equiv \langle E_{n_z,n_r,\upsilon}(q+\phi)-E_{n_z,n_r,\upsilon}(q)\rangle \approx  -2\hbar J_{n_z}^{T_r}(\cos(q+\phi)-\cos(q))$, where $\langle \bullet\rangle$ denotes radial thermal averaging, and $n_r$ and $\upsilon$ the radial and azimuthal harmonic oscillator mode numbers. Thus one can replace the effective Rabi coupling $\Omega^{\pm{\rm eff}}_{n_z,\ell,q,\upsilon}$ for the carrier ($\ell=0$),
$\Omega^{\pm {\rm eff}}_{n_z,0,q,n_r,\upsilon}\to \Omega^{\pm{\rm eff}}_{n_z}(q)=\sqrt{\Omega_{n_z,0,q}^2+\left(\pm \delta+4J_{n_z}^{T_r}\sin(\phi/2) \sin(\phi/2+q)
\right)^2}$, where $J_{n_z}^{T_r}$ is the thermally averaged tunneling rate, and $\Omega_{n_z,\ell,q}$ is the Rabi coupling for the $\ell^{\mathrm{th}}$ axial sideband. Even when the tight-binding approximation no longer holds, we can reproduce the VHS splittings by matching the thermally averaged bandwidth, computed from the analytic expression for the perturbative energies given above, to the tight-binding expression, $\max_q\langle \Delta E_{n_z,n_r,\upsilon}(q)\rangle-\min_q\langle \Delta E_{n_z,n_r,\upsilon}(q)\rangle=8\hbar J_{n_z}^{T_r}\sin(\phi/2)$.  This expression, together with the approximation that all atoms are initially in the $\ell_0=0,1$ band and that $\hbar J_{\ell_0}^{T_r}\ll k_B T_{z} $, is used to fit the carrier lineshapes shown in Fig.~2  by convolving the resulting Rabi lineshape with the joint transition density of states $\mathcal{D}_{\ell_0,T_r}$:
\begin{eqnarray}
\bar{P}_{0}^\pm &\approx&  \frac{1}{4\pi }\int_{-\pi}^{\pi} dq  \left|\frac{\Omega_{ \ell_0,0,0}}{\Omega^{\pm \rm eff}_{\ell_0}(q)}\right|^2= \frac{1}{2 } \int_{-|4J_{\ell_0}^{T_r}\sin(\phi/2))|}^{|4J_{\ell_0}^{T_r}\sin(\phi/2)|} d\epsilon \mathcal{D}_{\ell_0,T_r}(\epsilon)\frac{\Omega_{ \ell_0,0,0}^2}{\Omega_{\ell_0,0,0}^2+\left(\pm \delta-\epsilon\right)^2}
\end{eqnarray}
where
\begin{align}
\mathcal{D}_{\ell_0,T_r}(\epsilon)=\left\{\begin{array}{cc } \frac{1}{\sqrt{(4J_{\ell_0}^{T_r}\sin(\phi/2))^2-\epsilon^2}}\, & -|4J_{\ell_0}^{T_r}\sin(\phi/2)|\le \epsilon\le |4J_{\ell_0}^{T_r}\sin(\phi/2)|\, \\ 0 & \mbox{otherwise}\end{array}\right.\, .
\end{align} 
When fitting this function to the split line data of the type shown in Fig.~2a,b to extract the VHS splitting plotted in Fig.~2c, the only free parameters used are a single parameter for the thermally averaged tunnel coupling $J$, and an additional parameter for the amplitude of the split line. The amplitude is only required as a free parameter when a pulse longer than a single $\pi$-pulse is used to increase the excitation fraction. The no free-parameters theory predictions plotted in Fig.~2c were generated by applying the perturbative model to measured spectra of the motional sidebands at each axial trapping potential. As mentioned above, the radial temperatures used for the perturbative model varied between $7-9$$~\mu$K.  

\begin{figure}
\centering
\includegraphics[width=0.99\columnwidth]{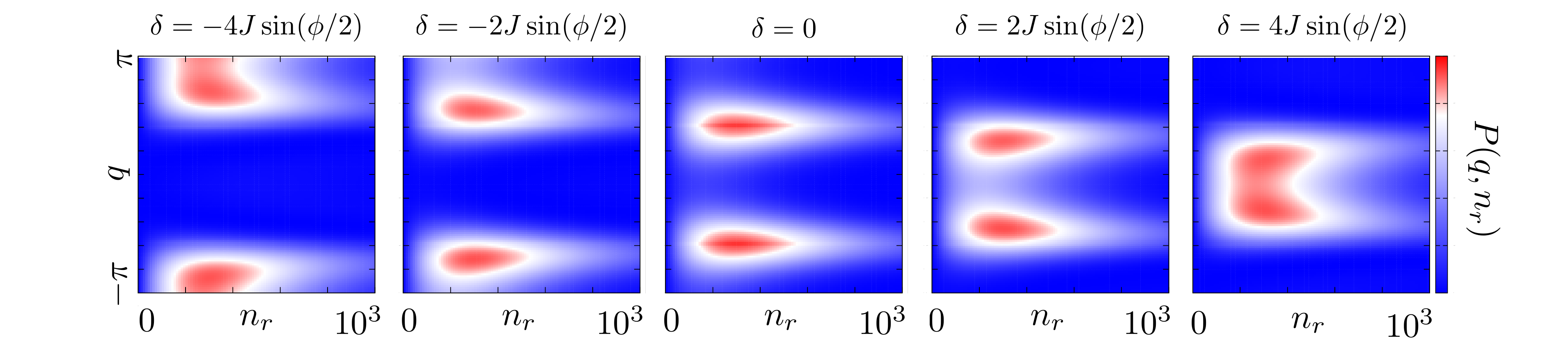}
\caption{Theoretical probability distribution of quasimomentum and radial quantum number $\mathbf{n_r}$ of $|e\rangle$-state excitations resulting from exciting a thermal distribution in $|g\rangle$ with a 50~ms $\pi$-pulse. The lattice depth is $U_z\sim 16~E_R$, resulting in a tunneling rate of $J \sim17$~Hz. The distribution of atoms among $q$ is broadened due to the fact that the Rabi frequency and tunneling rates are comparable, but can be made narrower by decreasing the Rabi frequency.}\label{fig:dist}
\end{figure}

\noindent\textbf{$\bullet$ Chiral Bloch vector dynamics:} To model the chiral Bloch vector dynamics, the Rabi oscillations measured after preparing the atoms within a window of quasimomenta centered around $q^*$ and width $\Delta q$ are expressed as,

\begin{align}
\bar{P}_{0}^+(q^*)= \frac{1 }{\Delta q} \sum_{q\epsilon \Delta q} \left|\frac{\Omega_{ 0,0,0}}{\Omega^{+ \rm eff}_{0}(q^*+\Delta q)}\right|^2 \sin^2\left[\frac{t}{2}\Omega^{+ \rm eff}_{0}(q^*+\Delta q)\right].
\end{align}
To model the window, we simulate the atom preparation using a $\pi$-pulse ($\Omega=2\pi\times10$~Hz) with five different initial pulse detunings $\delta^{*}$. Because the Rabi frequency and tunneling rates are comparable, the window is relatively broad, and this results in dephasing of the quasimomenta-dependent Rabi oscillations shown in Fig.~4d when a second, stronger Rabi pulse is applied. Fig.~ED\ref{fig:dist} shows the expected quasimomentum distributions. Additionally, for the theory lines presented in Fig.~4, we set
 $\Delta=(4J_{\ell_0}^{T_r}\sin(\phi/2))= 2\pi\times$ 67$ {\rm~s}^{-1}$ and $\Delta q\sim L/2$. 
 
%
\pagebreak

\spacing{1.5}

\noindent\textbf{References}
\bibliographystyle{naturemagmethods}
\bibliography{SOCmethods}

\pagebreak

